# Deterministic site exchange cellular automata models for the spread of diseases in human settlements.

Ricardo Mansilla[1,2], José L. Gutierrez[3]

[1)] Department of Complex Systems

Physical Institute, National University of Mexico

[2)] Faculty of Mathematics and Computer Science

University of Havana, Cuba

[3)] Agroecological Program, Chapingo University, Chap., Mexico.

## Abstract

A cellular automata model that describes as limit cases of his parameters the spread of contagious diseases modeled by systems of ordinary or partial differential equations is developed. Periodic features of the behavior of human settlement are considered. The model is built taking into account the range of motion of the elements of population. For small (large) values of this range, the behaviors described by partial (ordinary) differential equation models are reproduced. Emphasis is done in the study of those scenarios in which the above mentioned equations fail to describe. Some interesting behaviors in these cases are reported.

## 1 Introduction.

One of the first tasks that faces a theoretician who wants to interpret the time evolution of a complex system is the construction of a model. In real systems, many of their features are likely relevant, but not all of them, however, are included in the model. Although this lack, such simplified description should often be very helpful in developing the necessary intuition for the understanding of the behavior of real complex systems.

Most models in population dynamics, spread of diseases, rumors and news are formulated in terms of differential equations, both partial and ordinary. The difference between the types of equations used is strongly related with the assumptions made about the way members of population move in the spatial region they belong to. If the model assumes a homogeneous mixing of the different classes of individuals (infected, susceptible, removed, etc.) then, systems of ordinary differential equations are used. If short range character in the motion is assumed, that is, if the mean length of the motion of members of population is small respect to the size of the spatial ensemble where they live, then the spread of some diseases behave as diffusion processes and partial differential equations are used in their modeling. In both cases the assumption of random motion is made for the individuals.

Two objections could be made about these assumptions. First, the diffusive or perfect mixing hypothesis is not quit fulfilled always. Second, the motion of individuals of some species (human among them) is not random as assumed in the hypothesis of the above mentioned models. At least for human population it is far to be true. A lot of people go to school or to work and later go back home daily, therefore periodic motion seems more praiseworthy that random. Many other species (as foxes and other mammalians) have also motion routines.

Besides, few have been said about those transmission processes that do not fit very well in the mentioned paradigms. The aim of this work is develop a cellular automata model, which contains as limit cases the diffusive and perfect mixed behaviors mentioned above. This model depends on two parameters, the mean path length motion of individuals $\lambda$ and a measure of the strength of the contagion process $p$. Tuning the parameter $\lambda$ we obtain the diffusive or the perfect mixed behaviors and also describe those scenarios hard to characterize with differential equations. The structure of the paper is as follows. In Sec. 1 we develop the cellular automata model. In Sec. 2 discuss the relationship of $\lambda$ and the limit cases described by differential equation models. In Sec. 3 discuss the result of simulations and Sec. 4 is for conclusions.

## 2 The model.

Cellular automata provide simple models for a variety of complex systems containing a large number of identical elements with local interactions. A cellular automata consists of a lattice with a discrete variable at each site, evolving at discrete time steps. At a given time, the value of the variable at one site is determined by the values of the variables at the neighboring sites. The neighborhood could include the site itself. The evolution rule is synchronous, that is, all sites are updated simultaneously. Cellular automata are therefore, discrete (in space and time) dynamical systems. For a review of the main properties see for examples Gutowitz, 1990, Manneville *et al.*, 1989 and Wolfram, 1983.

Site exchange cellular automata are automata networks whose rules consist of two subrules. The first, applied synchronously, is a local rule inspired in the Conway's 'Game of Life'. It describes the local behavior of the transmission processes (contagion process, spread of news and rumors). We call this type of rules *contagion rules* although their validity is far beyond the scope of infective processes. The second, which has been sequentially applied, describes the motion of a fraction of individuals. We call this type of rules *transport rules*. These models have been extensively studied (Boccara and Cheong, 1992, Boccara and Cheong, 1993 and Boccara *et al.*, 1994) in the last years.

Let $Z$ be the set of the integer numbers and $\Lambda \subseteq Z^2$ be a lattice. The set $\Lambda$ represents the spatial environment where the population lives. At a time step $t$ a site of $\Lambda$ is either empty or occupied (representing an individual in some subclass of the population). The way the transport subrules have been used is as follows (see Boccara and Cheong, 1992, Boccara and Cheong, 1993 and Boccara *et al.*, 1994): Each time step, an occupied site selected at random is swapped with another site (empty or occupied) also selected at random. This operation is repeated $mc(t)N$ times, where $N$ is the total number of sites, $c(t)$ the density of nonzero sites at time $t$ and $m$ is a parameter called the *degree of mixing*. It is important to note the stochastic character of the process. These rules fail to explain why in a human population, with rigid motion schedules dictated by the day routine the spread of epidemics behave under certain conditions as perfect mixed or perfect diffusive.

In order to fill this gap we device another type of transport subrule. Let $\Lambda \subseteq Z^2$ whose vertices are occupied by members of population. Let $\Omega = \{0, 1, \ldots, p\}^{\Lambda}$ be the set of element of the form $(a_{(i,j)})_{(i,j)\in\Lambda}$ where $a \in \{0, 1, \ldots, p\}$ represents to which subclass they belong and $(i, j)$ is a position in the lattice. For example, in an epidemic process these subclasses could be empty, susceptible, infective, removed. The set $\Omega$ contains all the possible configurations over the lattice $\Lambda$. Hence we call it *configuration space.*

Let $\tau : \Omega \to \Omega$ be a function which satisfy the following conditions:

a) Let $(\tau(a_{(i,j)}))_{(i,j)\in\Lambda}$ be the image of the element $(a_{(i,j)})_{(i,j)\in\Lambda}$ under application $\tau$. Then if $a_{(i_0,j_0)} \neq 0$, then $\tau(a_{(i_0,j_0)}) = 0$.

b) For every $x_1, x_2 \in \Lambda, \quad x_1 \neq x_2$ such that $a_{x_1}, a_{x_2} \neq 0$, then $\tau(a_{x_1}) \neq \tau(a_{x_2})$.

The above statements deserve an explanation. Condition a) means that every non-empty element of the lattice can only be moved to an empty site. Condition b) means that two occupied sites cannot go to the same empty site. These are reasonable statements. We observe in Nature that elements of population by means of motion could be placed close together but never (except possibly in Manhattan Island) one over the other.

We call function $\tau$ a *transport rule*. We emphasize the synchronous character of this type of rule. In Fig. 1 we could see a schematic representation of a transport rule

Let $X \in \Omega$ be an element of $\Omega$. We denote by $O(X)$ the subset of $\Lambda$ of non-empty positions, i.e.: $O(X) = \{x \in \Lambda, a_x \neq 0\}$ and by $N(X)$ the number of elements of $O(X)$. We define the number:

$$\lambda = \frac{1}{N(X)} \sum_{x\in O(X)} \rho(x,\tau(x)) \qquad (1)$$

where: $\rho(x,\tau(x))$ is the Euclidean distance between the non-empty position $x$ and his destination by means of $\tau$. We call $\lambda$ the *mean path length* of motion of individuals.

Let define now the kind of local rules, which describe the transmission process (contagion, spread of rumors, etc.). Although we use concepts related with epidemic processes, the validity of definitions exposed below are far beyond this narrow framework.

Let take $\Omega = \{0, 1, 2\}^{\Lambda}$, where 0 means empty, 1 susceptible and 2 infective. Non-empty susceptible sites become infective by contact, i.e., a susceptible may become infective with a probability $p$ if and only if it is in the neighborhood of an infective. More precisely, during one time step, the probability that a susceptible having $n$ infected neighbors become infected is $1-(1-p)^n$. This hypothesis neglects latent periods, i.e., an infected susceptible become immediately infective. In Fig. 2 we could see a situation as described above. Some refinements on the transmission process just defined will be made along the paper.

Let $\varepsilon_p : \Omega \to \Omega$ be the function (which depend on probability $p$) representing the above mentioned transmission process. We call *contagion process* to the time trajectory of an element $X \in \Omega$ by the application $\pi = \tau^{-1} \circ \varepsilon \circ \tau \circ \varepsilon$:

$$\{\pi_n\}_{n \in N} = \{\pi^n(X)\}_{n \in N} \qquad (2)$$

Although this set up seems completely deterministic, it has a stochastic component. The function $\varepsilon_p$ could assign *in each realization* different elements $\varepsilon_p(X) \in \Omega$ to every element $X \in \Omega$. Hence the Eq. (2) should be understood as the set of realizations of the contagion process. The configuration $X$ is the initial conditions of the process. It encloses all the information at time $t = 0$. Models in ordinary or partial differential equations also use some information encoded in $X$ in form of initial and/or boundary conditions. Notice that the application $\pi$ represents the daily exposure of the elements of the population. They could be infected or not at their original sites ($\varepsilon$), later they move ($\tau$) to their destination, being exposed or transmitting the disease there ($\varepsilon$) and later go back to their original positions ($\tau^{-1}$). We oversimplify the process assuming that the only situations where the contagion takes place are at their original position or destination, i.e., homes, schools, jobs or any other social activities for human populations. In this case, exposure to epidemics during transportation (subway, bus, etc.) is neglected here. Similar assumptions can be made in animal diseases. Consider, for instance, rabies epidemic among the foxes. Rabies is a viral infection of the nervous central system. It is transmitted by contact and is invariably fatal. As stated in Kallen *et al.*, 1985 foxes acquire the disease mainly during hunting hours or at their dens.

## 3 Limit cases.

In this Section we show how for different values of the parameter $\lambda$ defined in Eq. (1), we could obtain the extreme behaviors of perfect mixing and perfect diffusion. All the simulations referenced in this Section were done with a lattice $\Lambda$ of 150x150 sites with half of them non-empty. The transmission process $\varepsilon_p$ will be as described in Sec. 2: there will be only susceptibles and infectives. More sophisticated scenarios are studied later. The total population is $N = 11250$. We always start the simulation with only one infective.

### 3.1 Large values of $\lambda$.

If the parameter $\lambda$ is large enough, then a perfect mixed behavior could be observed. In Fig. 3 is shown the pattern of infectives for several different values of $t$. The length of average path is $\lambda = 80$. Because the lattice size is 150, the value of $\lambda$ is extremely high and the perfect mixed behavior is easily observed. A classical differential equation model for this scenario is:

$$\begin{cases} \dfrac{dI(t)}{dt} = \alpha I(t)(11250 - I(t)) \\ I(0) = 1 \end{cases} \qquad (3)$$

where $I(t)$ represent the number of infectives in the instant $t$ and $\alpha$ is a constant related with the morbidity of the disease. The higher $\alpha$, the higher the growth rate of $I(t)$. We have observed a strong relationship between the constant $\alpha$ and the probability $p$ of the transmission process $\varepsilon_p$. In Fig. 4 is shown a graph of $\alpha$ versus $p$. Each point of the curve is obtained with 10 simulations with the same value of $p$. The time interval was $[0, 250]$. In each simulation the values of infectives $I_0, I_1, \ldots, I_{250}$ were introduced in a linear regression using Eq. (3) to obtain the value of $\alpha$. The process was repeated 10 times and the average is plotted.

## 3.2 Small values of $\lambda$.

With small values of $\lambda$ the cellular automata model behave as perfect diffusive. In Fig. 5 the pattern of infectives is shown for several different values of $t$. Notice the formation of a wave front, which grows until cover the entire lattice. This is in agreement with other results. Kallen *et al.*, 1985 added a diffusion term in the rate equation of infectives in the model proposed by Kermack and McKendrick (see Kermack and McKendrick, 1927 for details) in order to take into account the dispersion of rabid foxes. The new system of equations admitted traveling wave front solutions. Notice that if only exist susceptibles and infectives then, an elementary differential equation model for this scenario is:

$$\frac{\partial u(x,y,t)}{\partial t} = D\nabla^2 u(x,y,t) + \alpha u(x,y,t)(1 - u(x,y,t)) \qquad (4)$$

where $u(x, y, t)$ represent the density of infectives in the position $(x, y)$ in the instant $t$. The constant $D$ is the diffusion and the constant $\alpha$ is related with the morbidity of the disease. We have observed an interesting behavior of D with respect to $\lambda$. In Fig. 6 is shown a graph of $D$ versus $\lambda$. Each point of the curve is obtained with 10 simulations with the same value of $\lambda$. The procedure to construct the points is as follows: in order to find an approximation of $u(x, y, t)$, the lattice $\Lambda$ was divided in $15\times15$ squares of $10\times10$ sites. On each square the number of "infected" sites were counted and divided by 100. This is an approximation of $u(x_i, y_i, t)$ in the square centered at the point $(x_i, y_i)$ at time $t \in [0,250]$. All these values of $u$ are introduced in a nonlinear regression using Eq. (4) to obtain $D$ and $\alpha$. This process was repeated 10 times and the average values were plotted. Notice that as $\lambda$ grows the size of the error bars are larger. This fact is consequence of the breaking of the diffusive regimen. For values of $\lambda > 30$, the process is no longer diffusion.

The result of the Sec. 3.1 and Sec. 3.2 show that our cellular automata model could capture the main features of extreme cases described by differential equations, reported in the literature. In the following Section we study the properties of the "intermediate" zone for the values of $\lambda$.

## 4 The behavior for intermediate values of $\lambda$.

In this section we report our findings in the behavior of the site exchange cellular automata model for the values of $\lambda$ which are too large to yield diffusive behavior and too small to produce perfect mixing. That is what we call "intermediate zone" in the above section. Notice that for these values of parameter $\lambda$ the differential equation models fail to describe the behavior of the process.

We also add a new ingredient to our model, the length $d$ of the infective period, i.e., a susceptible, which becomes infective at time $t$, becomes again susceptible at time $t+d$. Let consider first a transmission process $\varepsilon_p$ as defined in Sec. 2, but assuming also that an infective become again susceptible $d$ time steps after she or he acquired the disease, i.e., we are assuming that the infection period has length $d$.

In Fig. 7 is shown the time series of infectives for $d=5,\ p=0.4,\ \lambda=45$. We report a scaling property for this time series. Let denote by $s_t = |I_{t+1} - I_t|$ the absolute value of the difference between two consecutive elements of the time series. Let $N(s)$ be the cumulative number of $s$, i.e., how many times the value $s$ appear in the $\{s_t\}$ series. We observe that:

$$N(s) \propto \frac{1}{s^{\delta}} \qquad (5)$$

It means that a small difference between two consecutive values in the time series of infective has a higher probability to appear than a larger one. We also found a strong correlation between $\delta$ and $p$. In Fig. 8 is shown the graph of $N(s)$ versus $s$ for *a realization* of the automata model. Notice that we plotted $N(s)$ for a realization, not for averaged values. In all cases studied the fitness of $N(s)$ to a power function was good. In the inbox plot of Fig. 8 are shown the averaged (over 10 simulations) values of $\delta$ versus the corresponding values of $p$. We confer a great practical value to the last result. It could allow the estimation of probability $p$ in real epidemic process having an accurate record of daily reported cases for several apparition of the disease.

We also studied periodic properties of the infectives time series. Our main tool was frequency domain analysis. Fourier spectra (see Percival and Walden, 1993). is widely used in time series analysis, because the visual representation in the frequency domain can more easily reveal patterns which are harder to discern in the primary data, for example, intricate periodical behavior. We use here Fourier transform of infectives time series to detect periodical features of that function. From now on, we call *power spectra* of infectives time series to the product of Fourier transform of that function by its complex conjugate:

$$\hat{S}(k) = \theta \left| \sum_{t=1}^{L} I(t) e^{-i2\pi \frac{k}{L} t} \right|^2 \tag{6}$$

where $\theta$ is a constant related with the sample frequency and $L$ is the number of data available for $I(t)$.

We studied the changes in the periodic behavior of $I(t)$ with respect to the order parameter $\mu = \frac{pd}{\lambda}$. We calculate the power spectrum of several infective time series with $\mu \in [0.06, 0.6]$. The results are presented in Fig. 9, where a behavior resembling period-doubling scenario is shown. It is well known that period doubling behavior is a possible scenario of chaotic processes. Evidence of this behavior in real contagion process have been reported in the literature.

## 5 Conclusions.

We have developed a cellular automata model for the spread of epidemics, rumors and news in a population of moving individuals. Our model depends on a parameter $\lambda$, which represents the means length of the motion of individuals in population. We reproduced with a suitable tuning of this parameter the limit cases of perfect mixing and perfect diffusion often described by systems of ordinary and partial differential equations respectively. We also could study those cases, which the above-mentioned models fail to describe. With these values of $\lambda$ and a suitable tuning of $p$ and $d$ we reported some evidence of period doubling behavior and other interesting properties. The authors are thankful to G. Cocho, O. Miramontes and P. Miramontes for their helpful comments. This work was partially (R.M.) supported by CONACYT, Mexico.

## 6 References.

## Figure captions.

**Fig. 1**: A schematic transport rule. Notice that some sites remain unchanged.

**Fig. 2**: A schematic contagion rule.

**Fig. 3**: Deployment of contagion process for $\lambda = 80$ for several values of $t$. Figure a) correspond to $t = 10$, b) to $t = 30$ and c) to $t = 50$.

**Fig. 4**: Rate of increase of infectives $\alpha$ versus the probability of contagion rule $p$.

**Fig. 5**: Deployment of contagion process for $\lambda = 15$ for several values of $t$. Figure a) correspond to $t = 10$, b) to $t = 30$ and c) to $t = 50$.

**Fig. 6**: Diffusion coefficient $D$ versus $\lambda$.

**Fig. 7**: Times series of infectives for $\lambda = 45$ and $d = 5$.

**Fig. 8**: $N(s)$ v.s. $s$. Notice the power law behavior. In the inbox is plotted the exponent of the power law fit with respect to $p$.

**Fig. 9**: Power spectrum of infectives time series for different values of $\mu$. Figure a) correspond to $\lambda = 0.0699$, b) to $\lambda = 0.3257$ and c) to $\lambda = 0.5896$. Notice the evidence of period doubling.

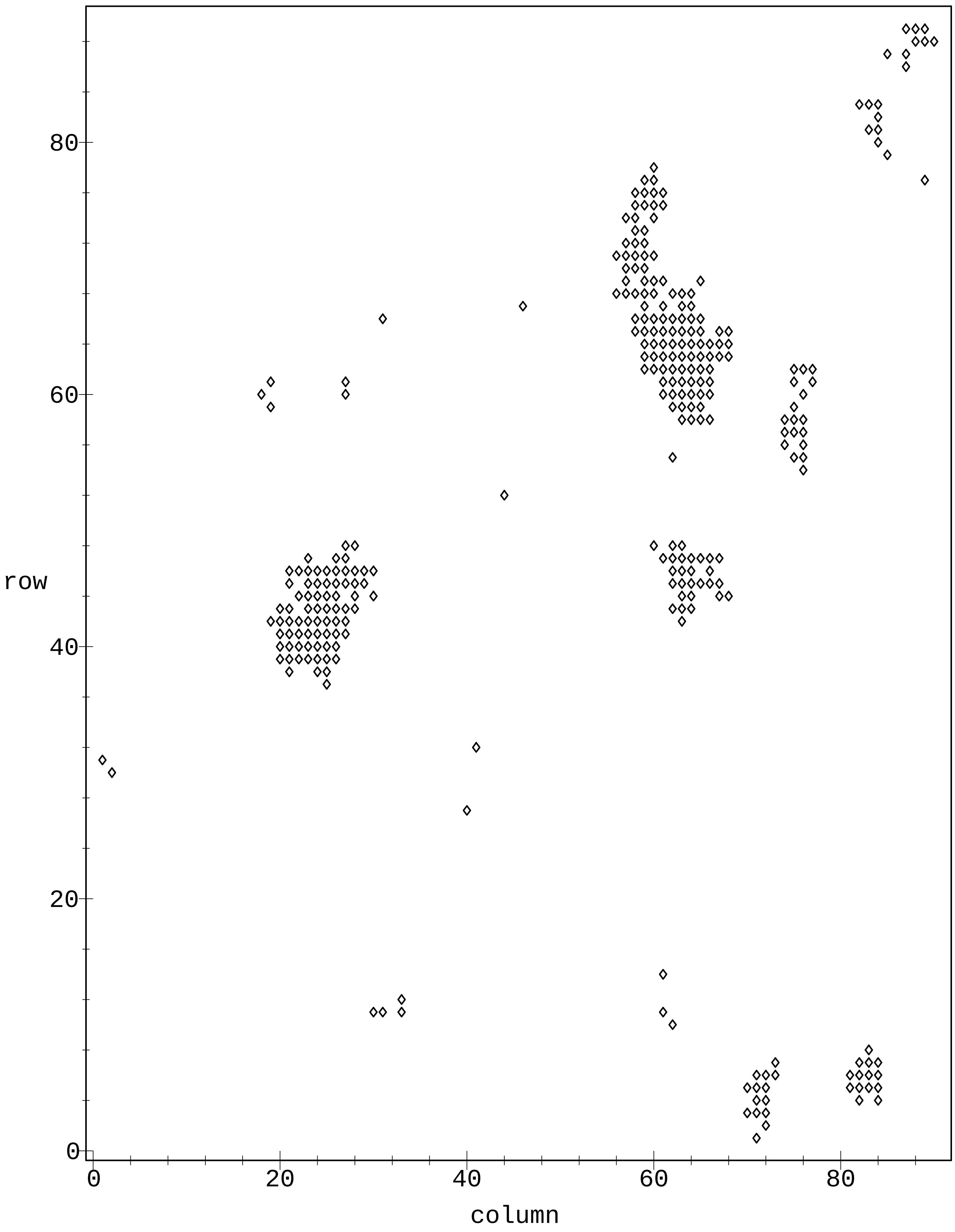

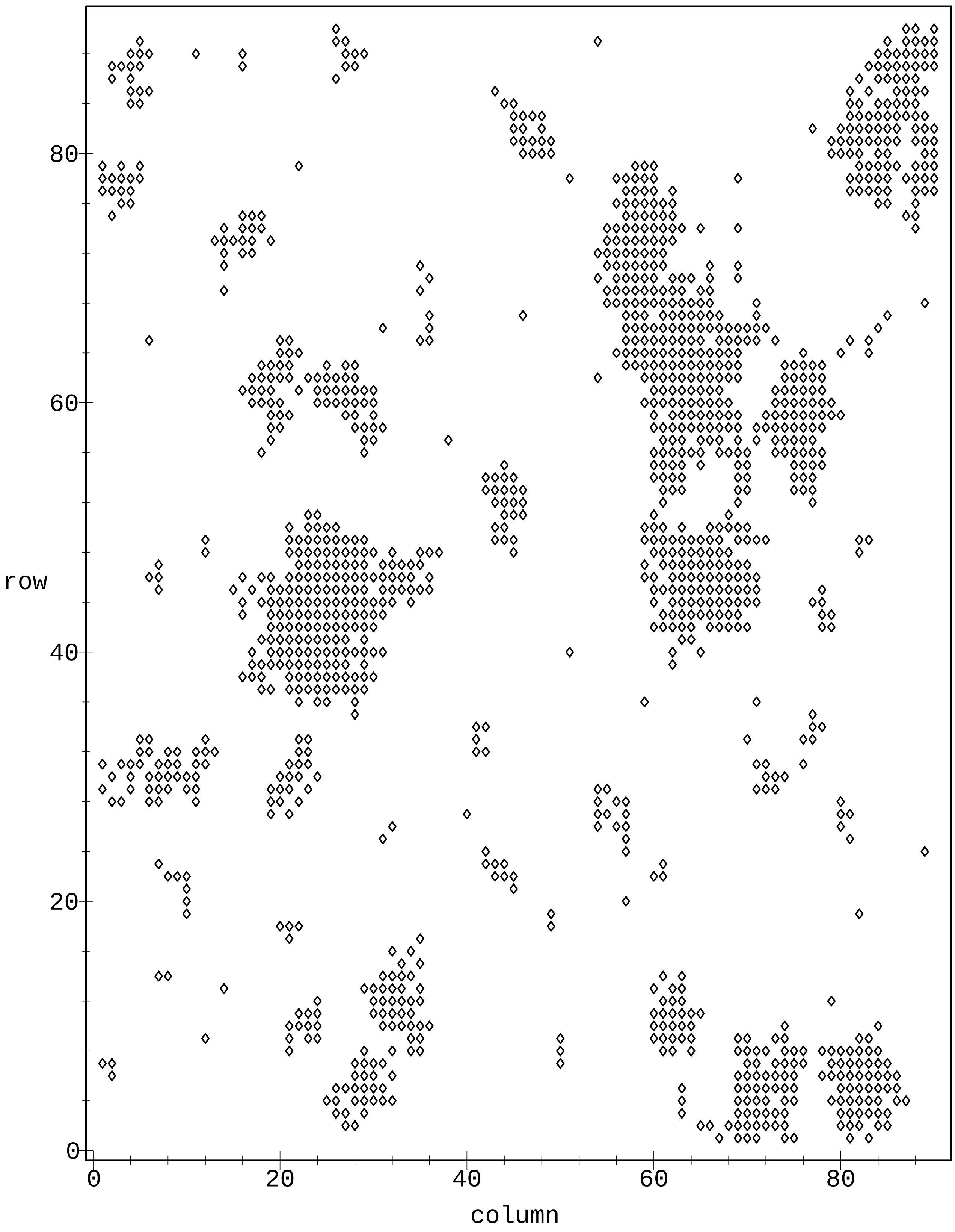

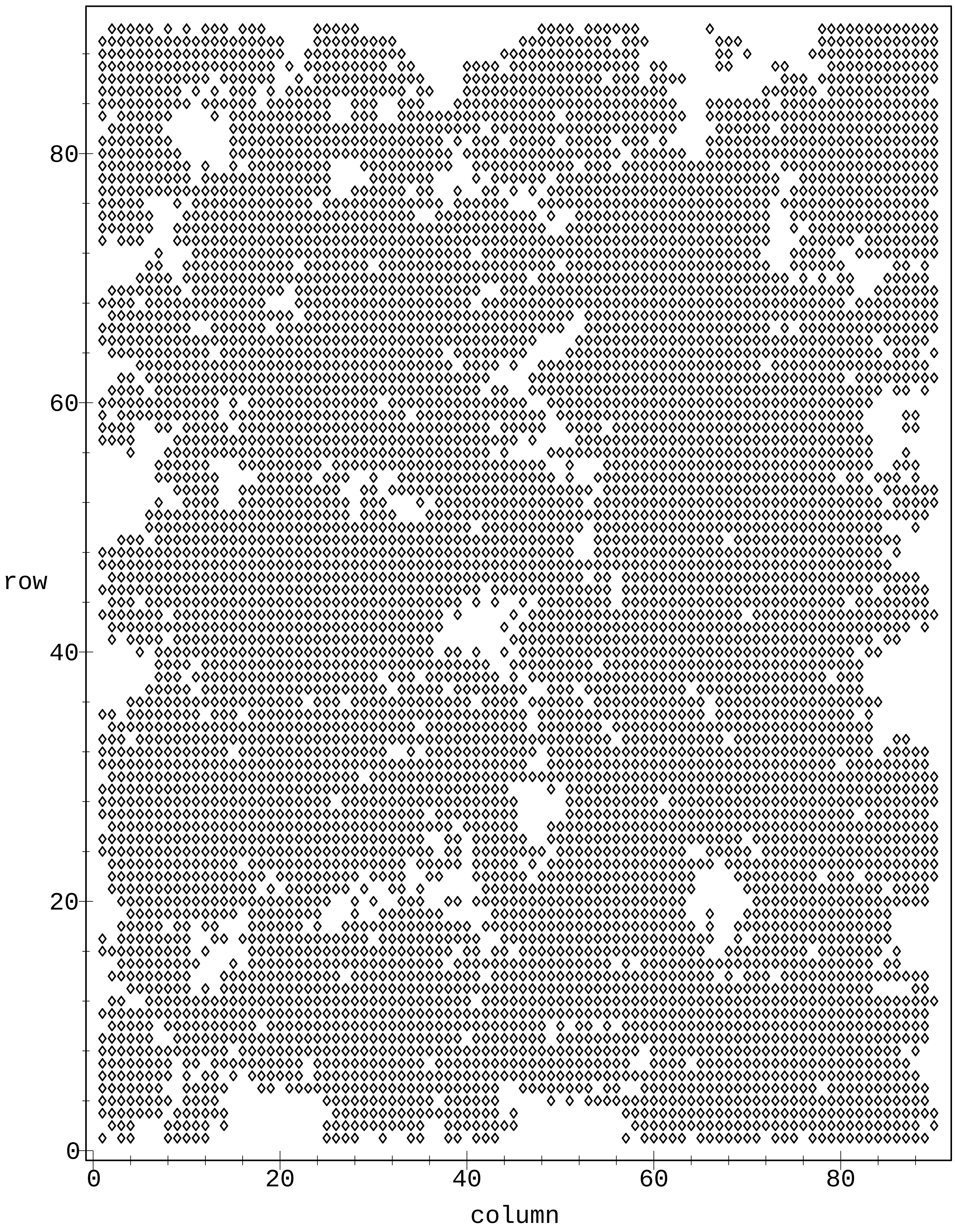

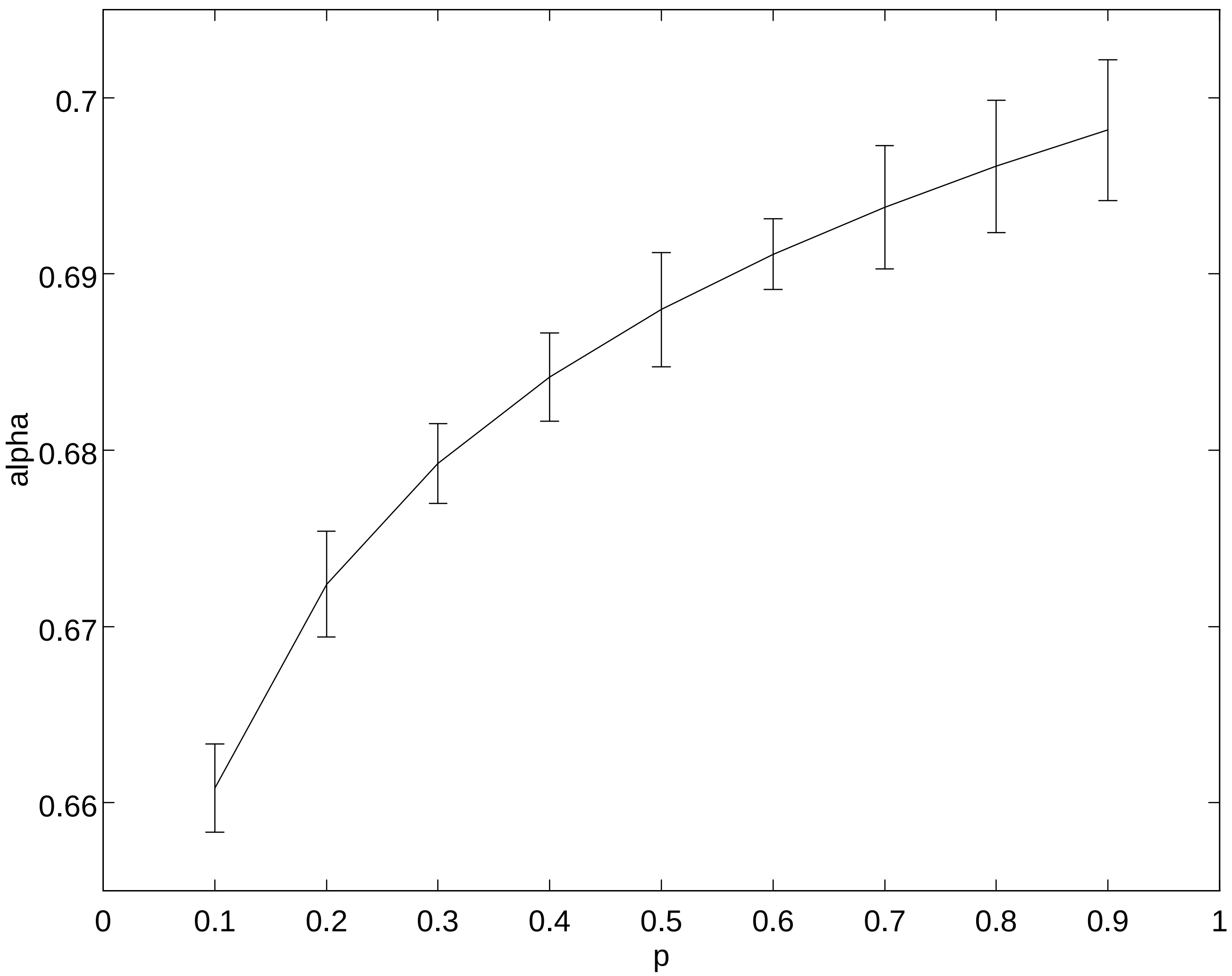

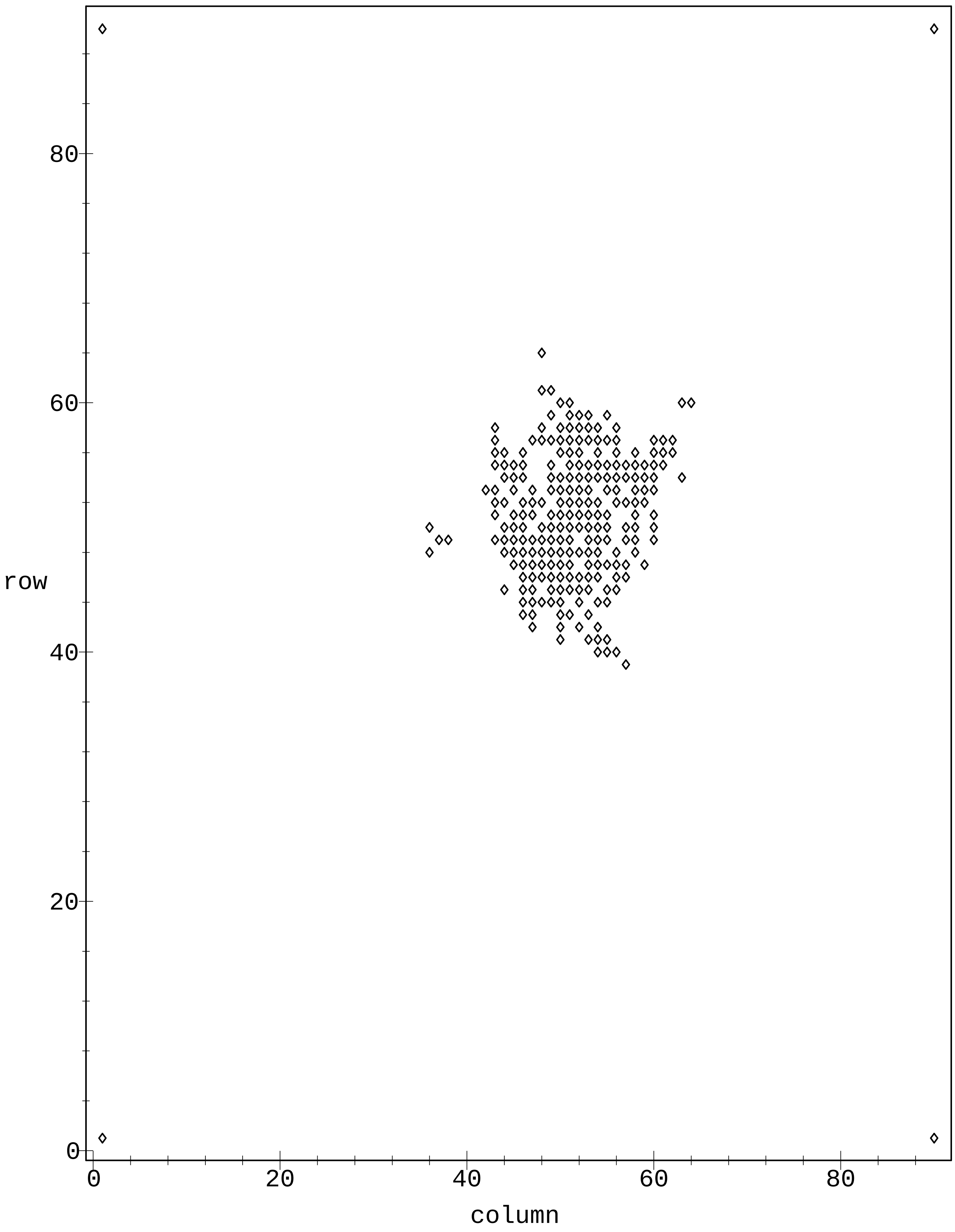

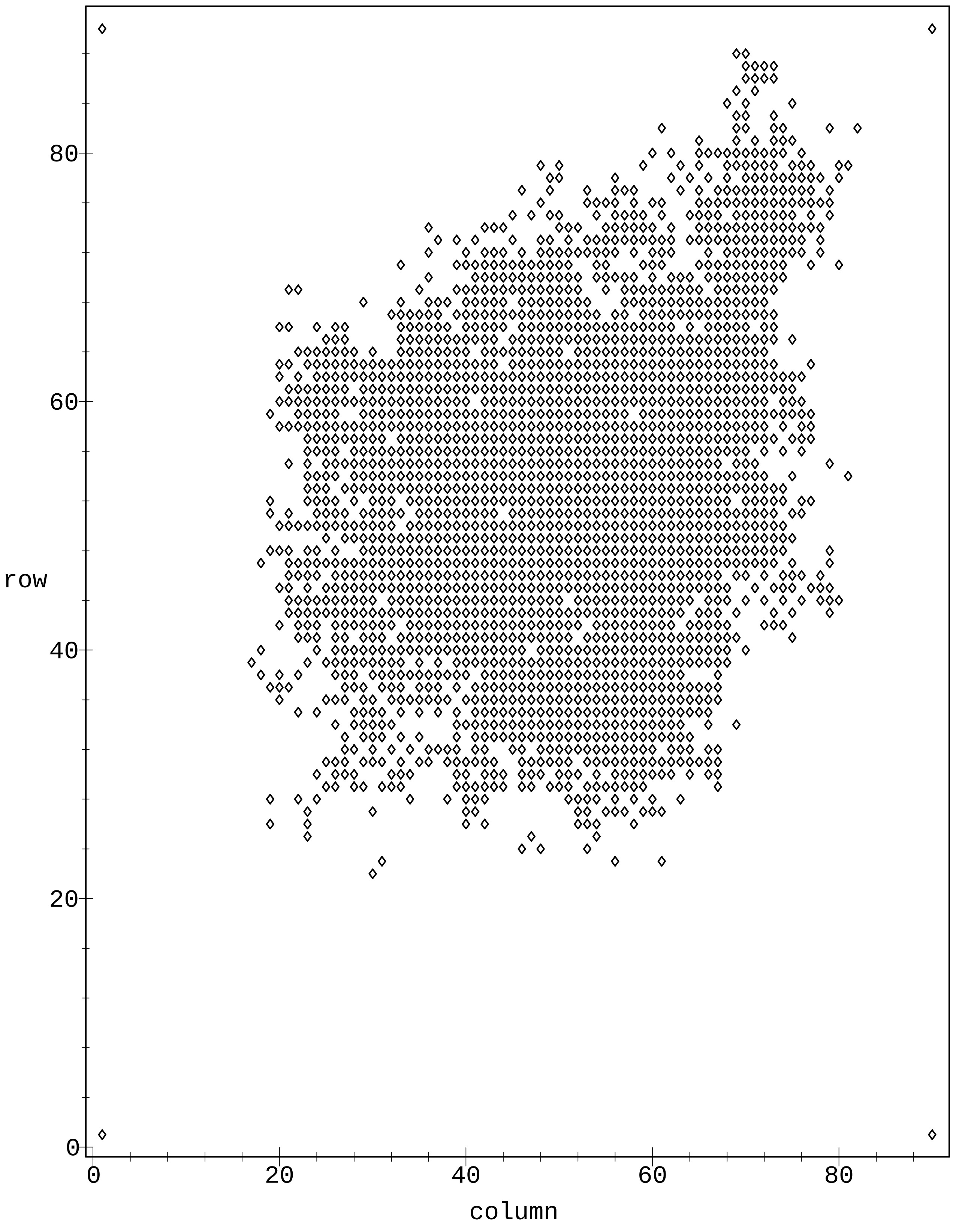

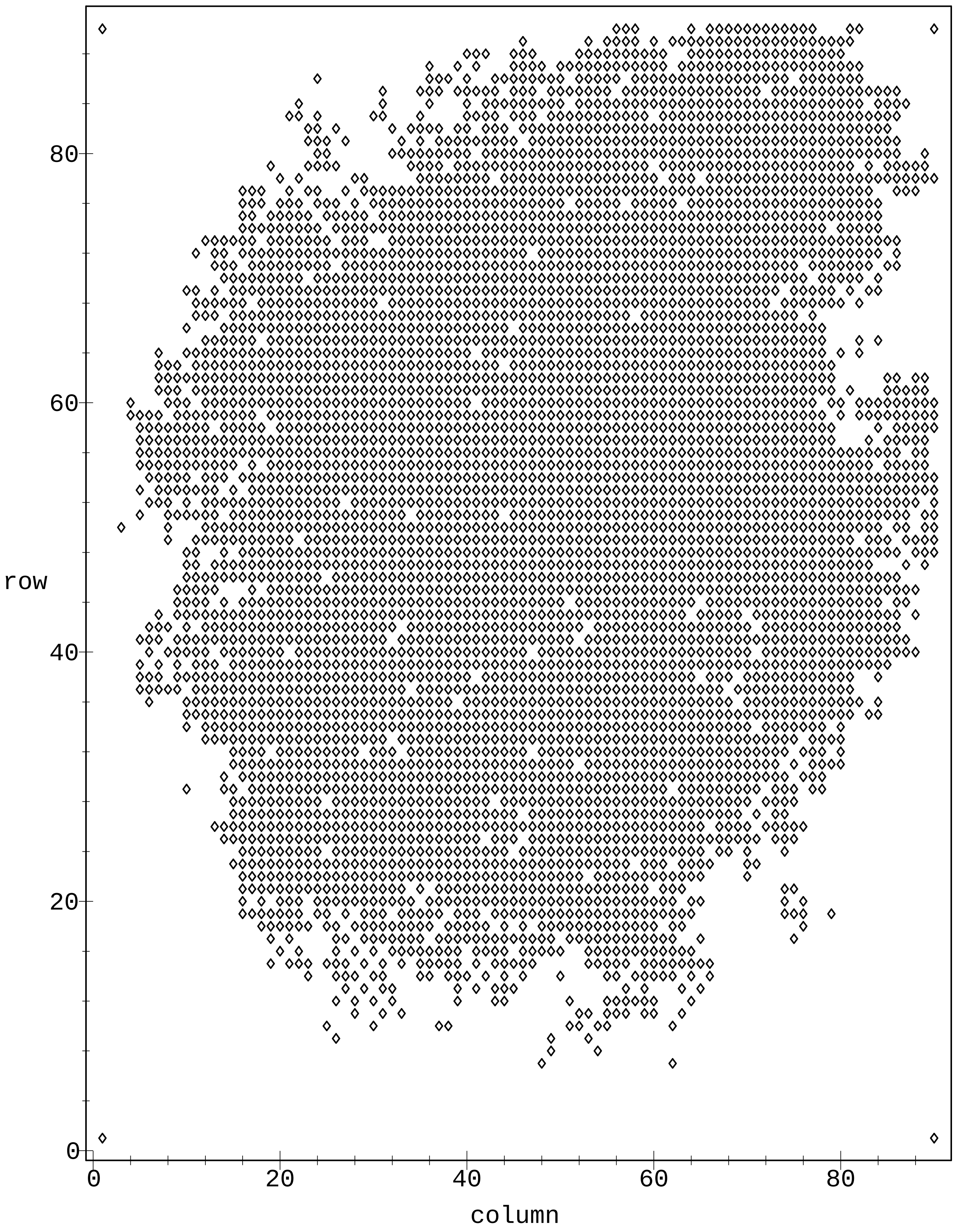

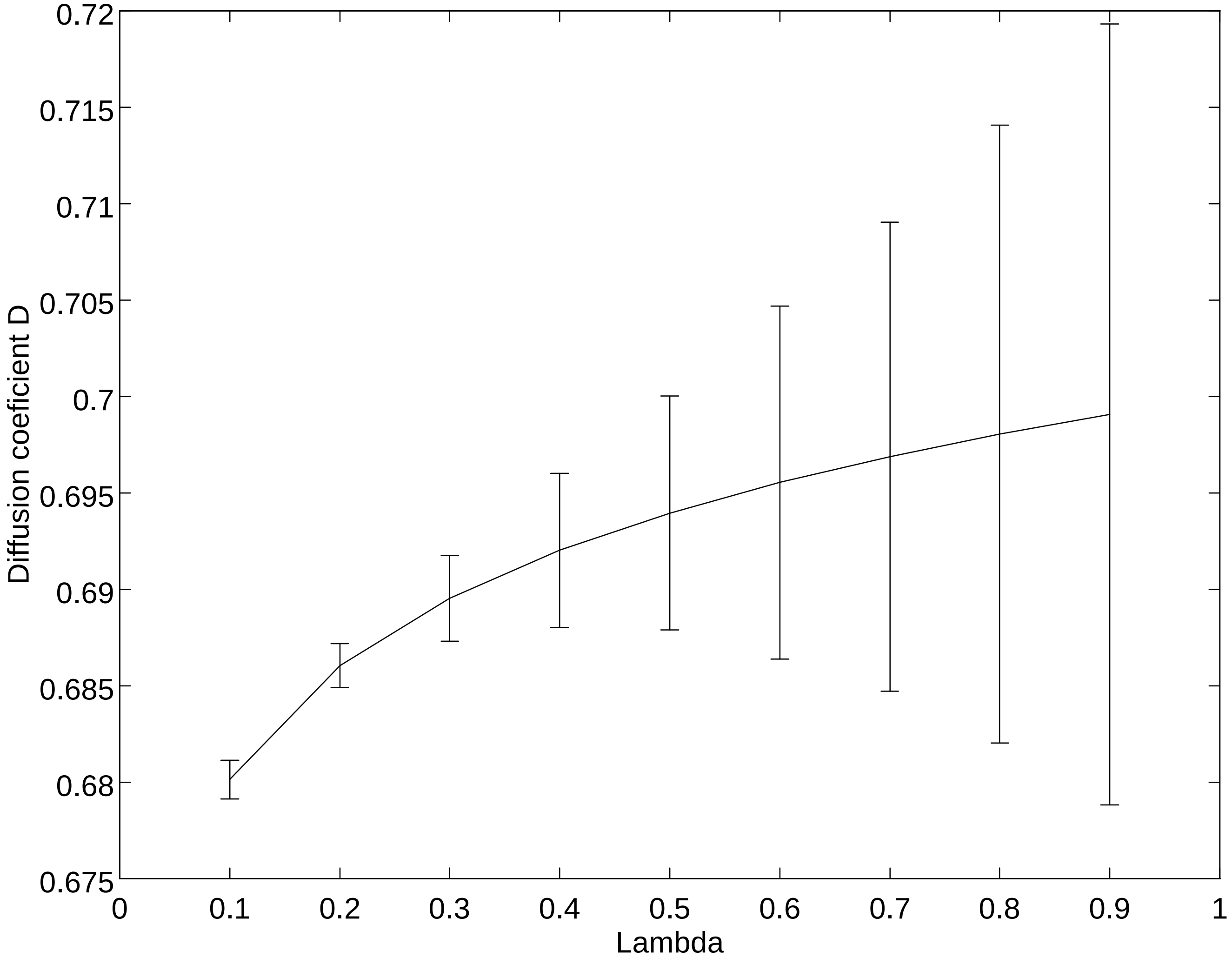

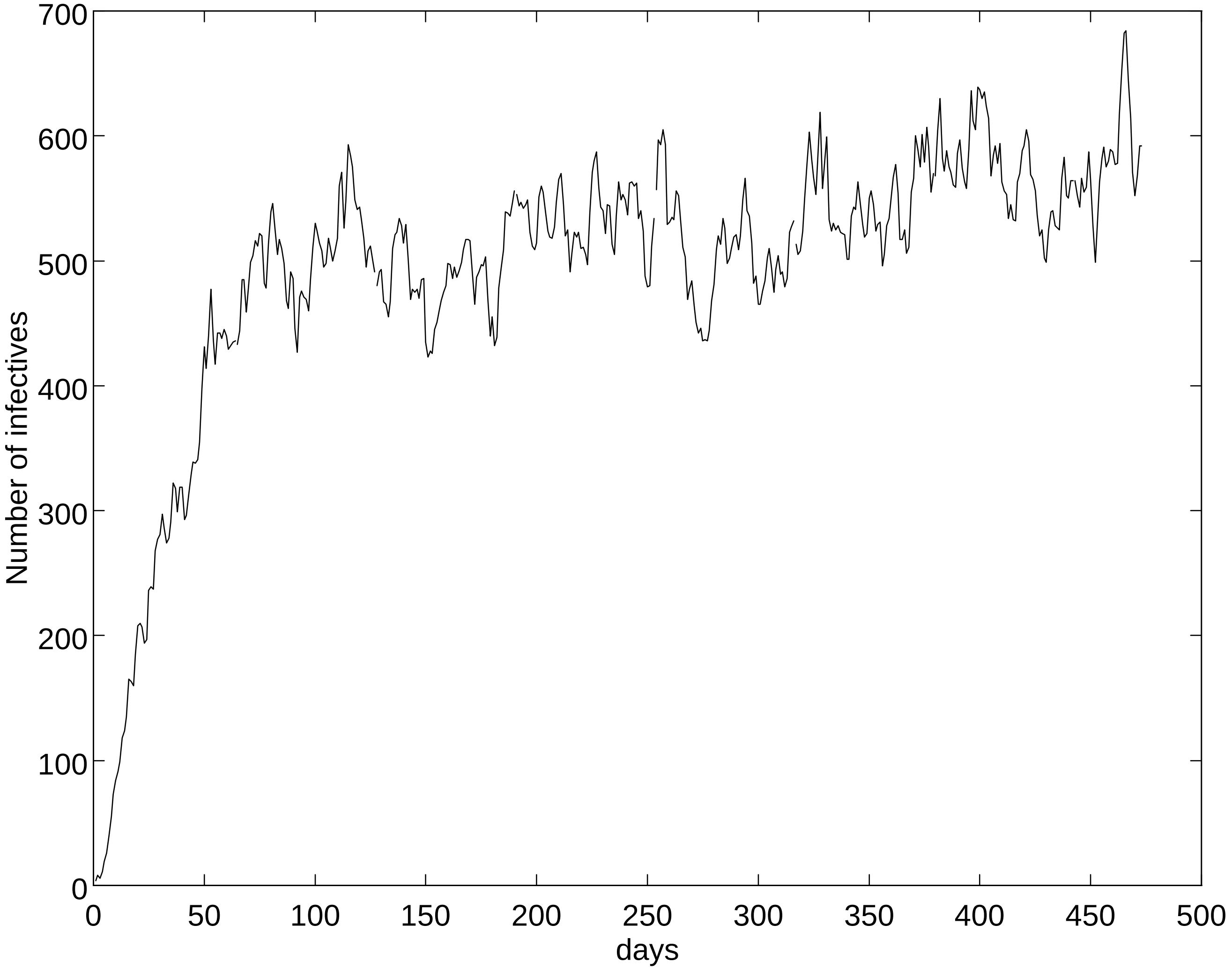

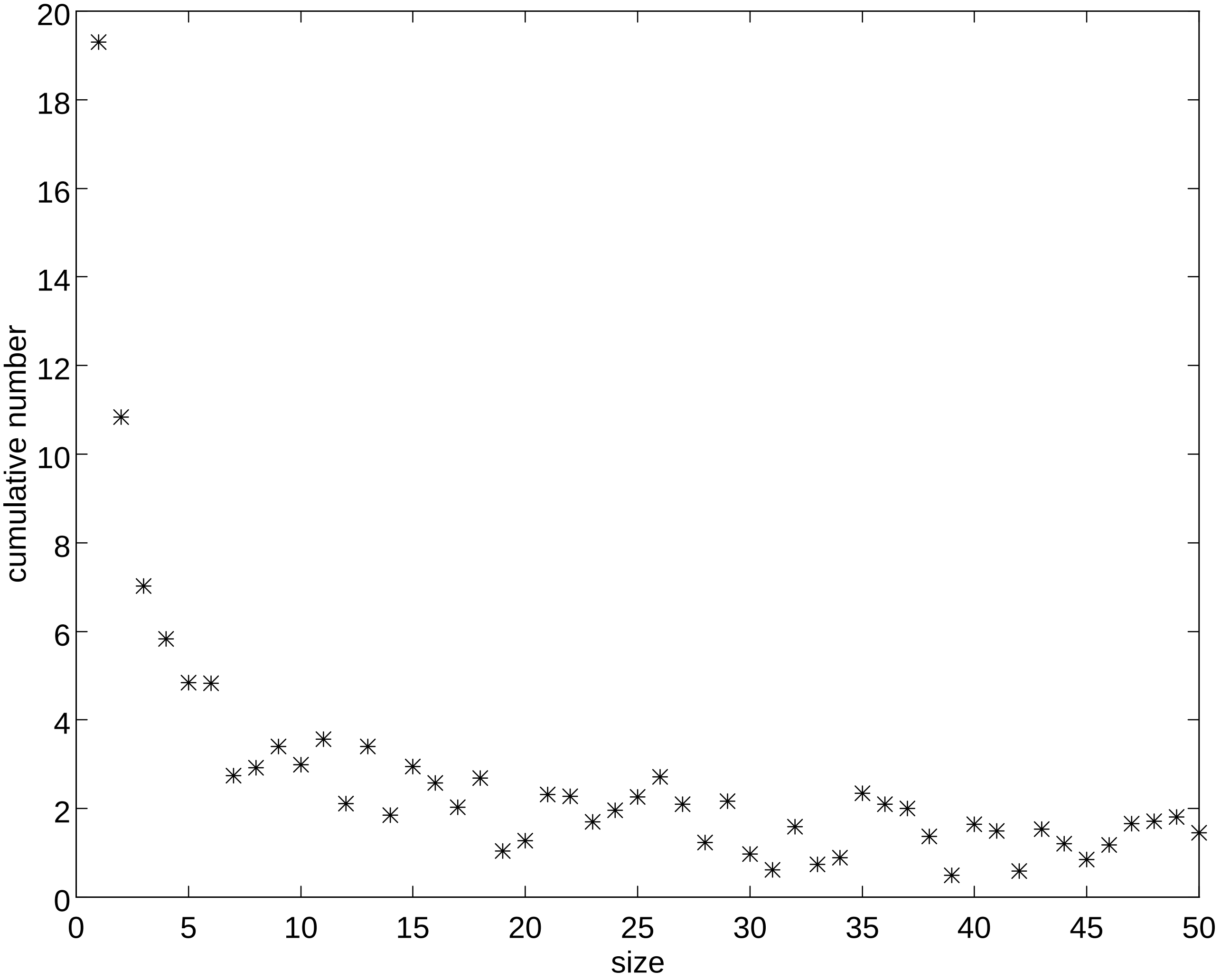

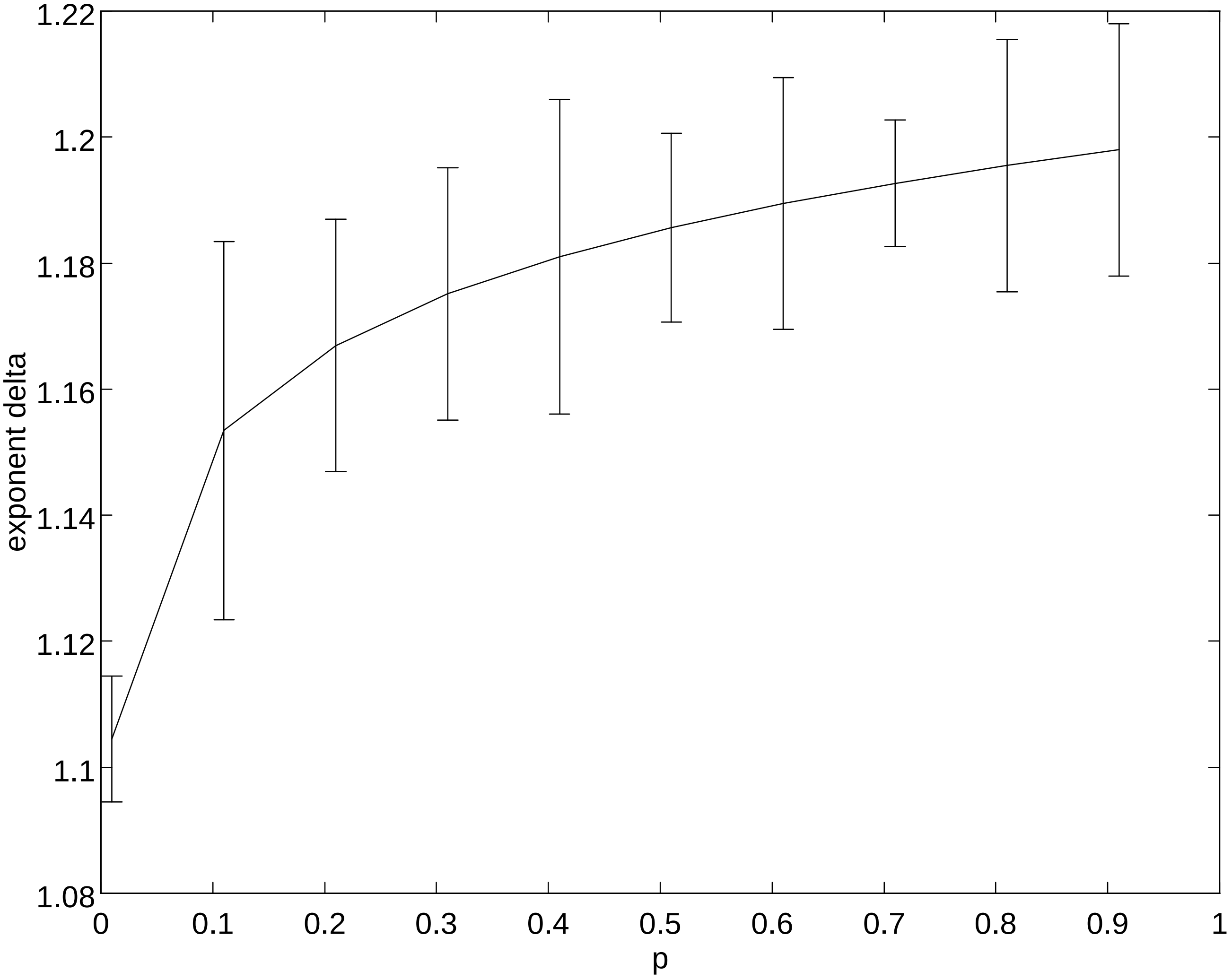

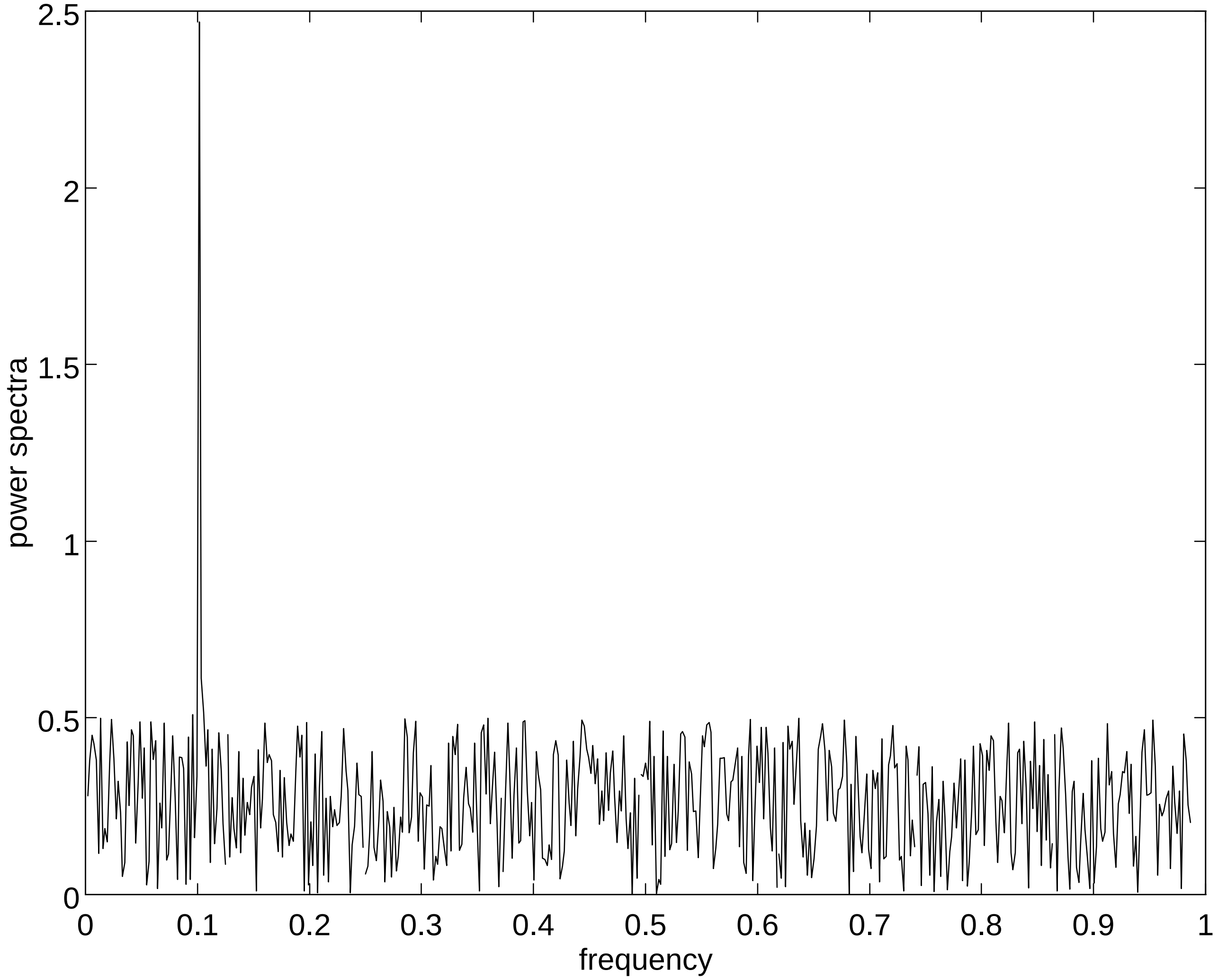

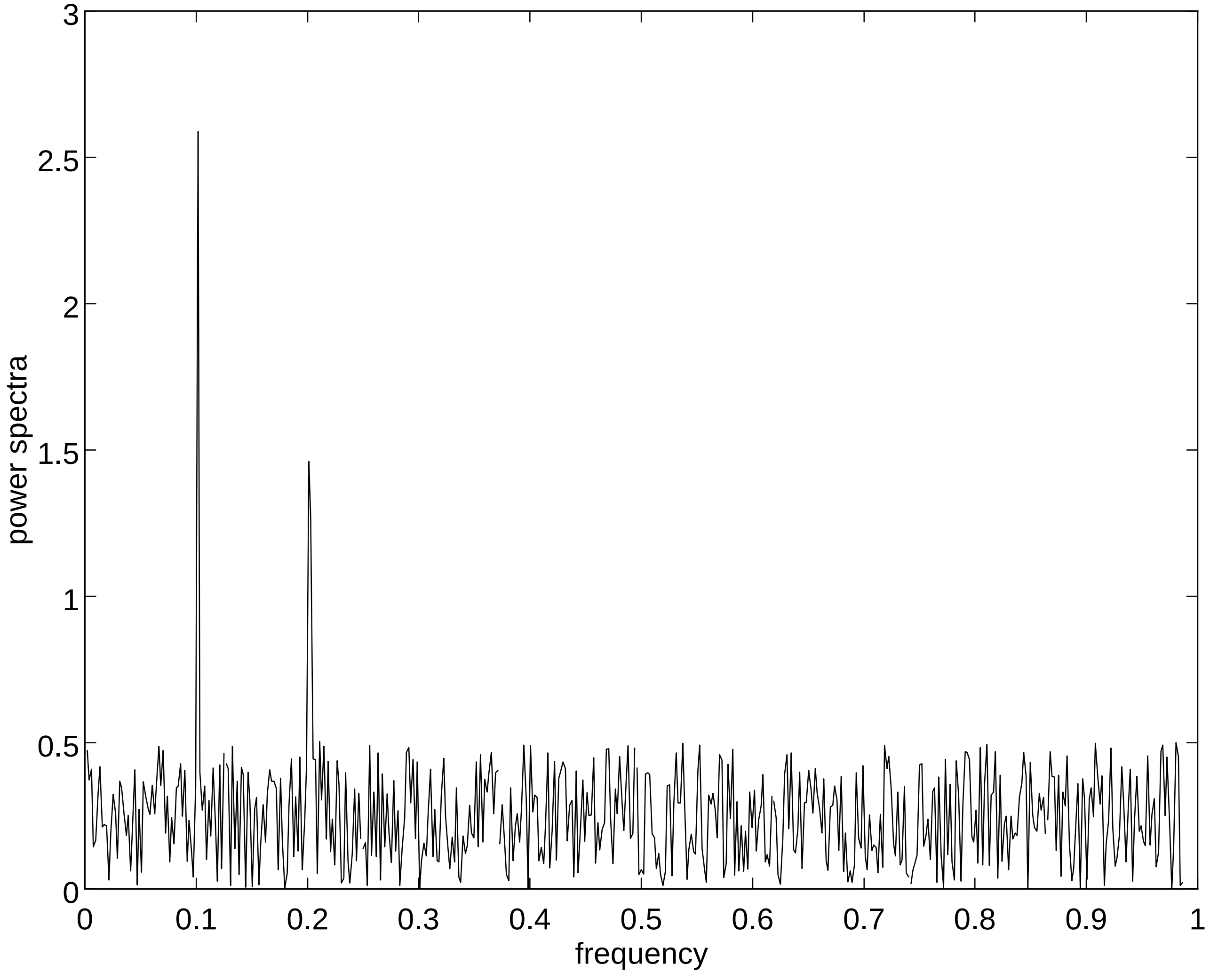

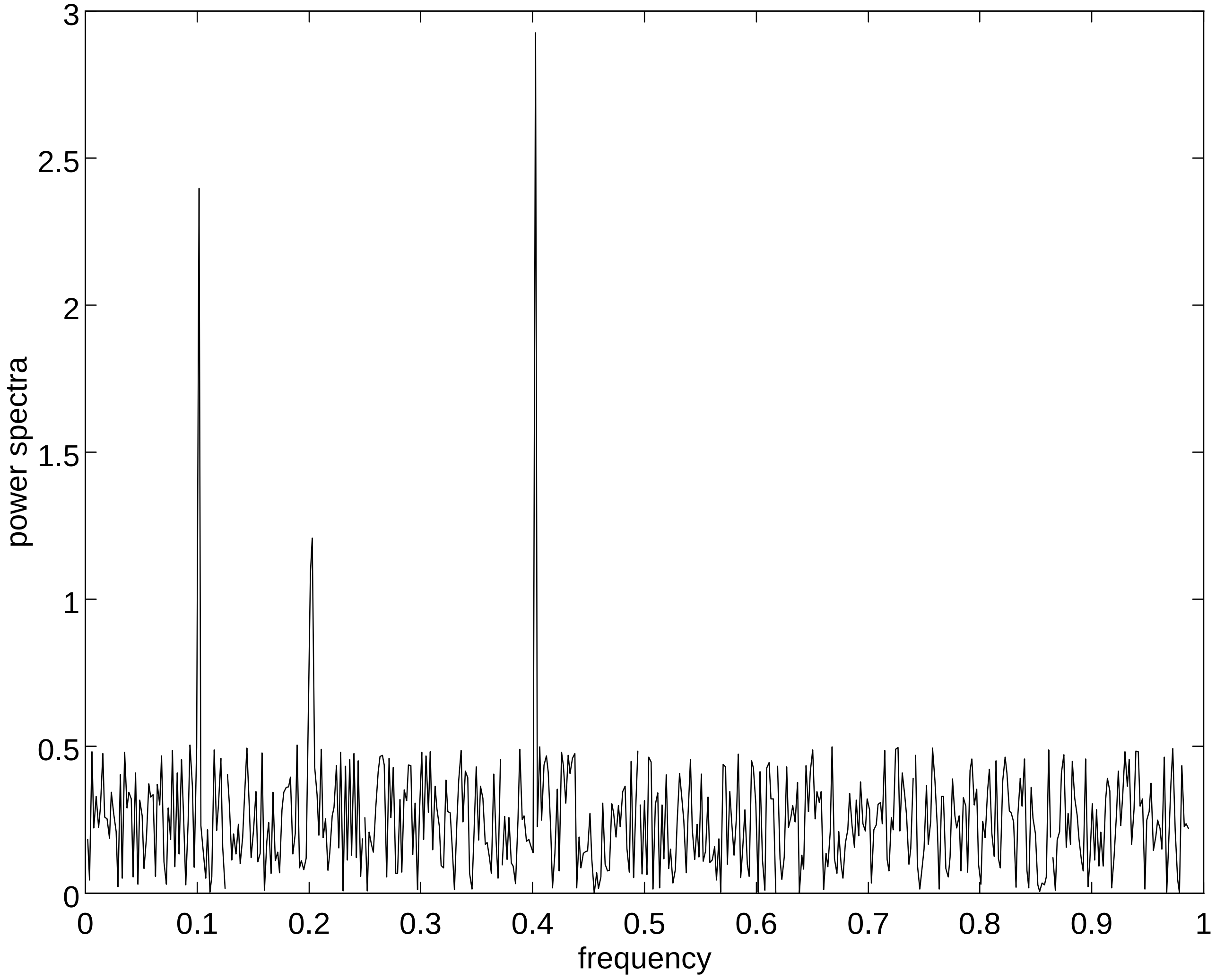

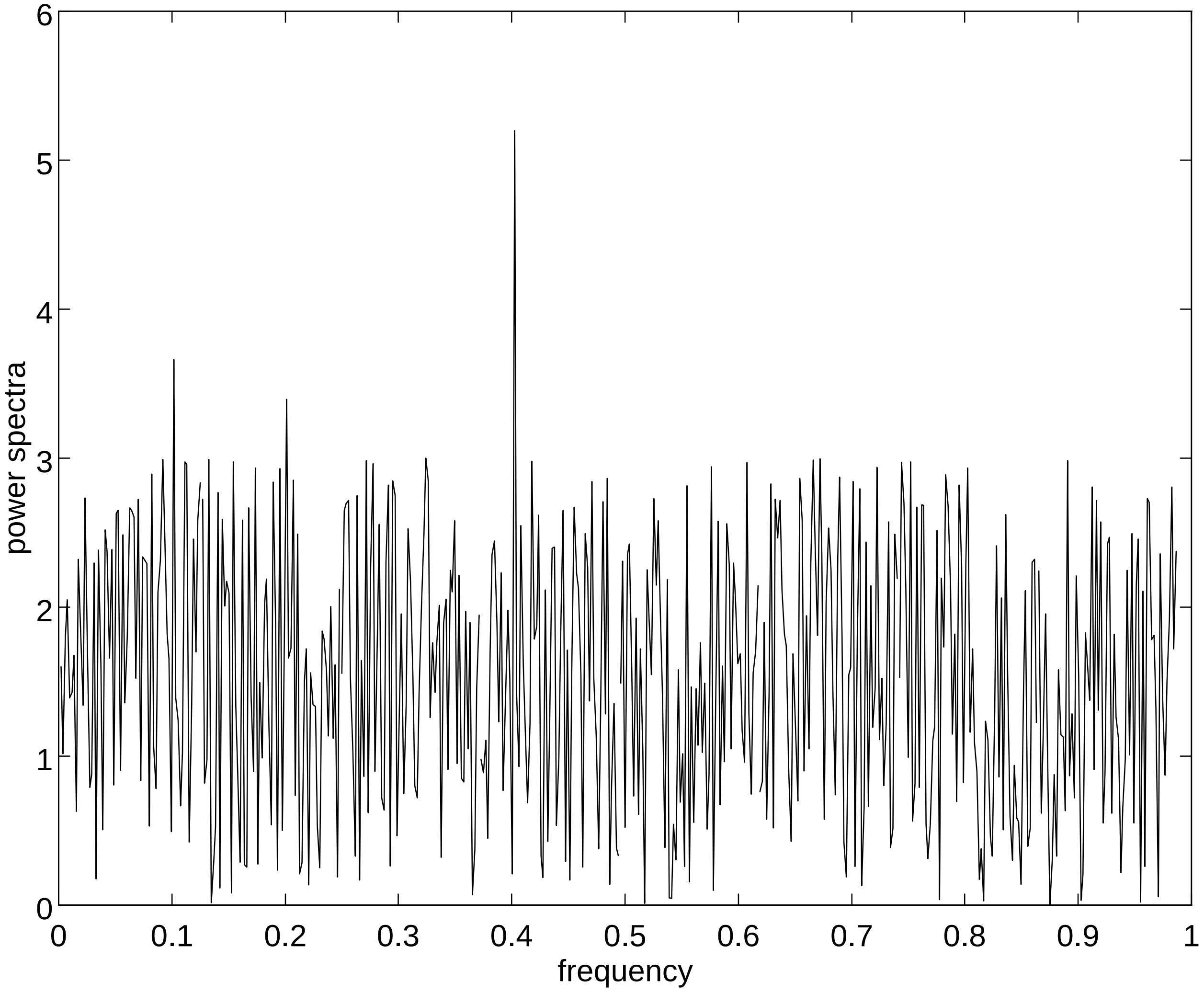